\documentclass[12pt]{article}
\usepackage{amsmath}
\def\a{\alpha}
\def\b{\beta}

\def\d{\delta}
\def\e{\epsilon}

\def\p{\psi}

\def\s{\sigma}

\def\be{\begin{equation}}
\def\ee{\end{equation}}
\def\arr{\begin{array}{rll}}
\def\ea{\end{array}}
\def\bea{\begin{eqnarray}}
\def\eea{\end{eqnarray}}

\def\N2{$N{=}2$}

\def\sfrac#1#2{{\textstyle\frac{#1}{#2}}}
\def\>{\rangle}
\def\<{\langle}
\def\+{\dagger}
\def\={\ =\ }

\begin{document}
\renewcommand{\thefootnote}{\fnsymbol{footnote}}
\begin{titlepage}
\setcounter{page}{0}
\begin{flushright}
ITP--UH--14/16\\
\end{flushright}
\vskip 2.0 cm

\begin{center}
{\LARGE\bf Superconformal SU($1,1|n$) mechanics}\\
\vskip 1.5cm
$
\textrm{\LARGE Anton Galajinsky\ }^{a} \quad \textrm{\Large and} \quad
\textrm{\LARGE Olaf Lechtenfeld\ }^{b}
$
\vskip 1cm
${}^{a}$ {\it
Laboratory of Mathematical Physics, Tomsk Polytechnic University, \\
634050 Tomsk, Lenin Ave. 30, Russian Federation} \\[4pt]
{Email: galajin@tpu.ru}
\vskip 0.5cm
${}^{b}$ {\it
Institut f\"ur Theoretische Physik und Riemann Center for Geometry and Physics, \\
Leibniz Universit\"at Hannover, Appelstrasse 2, 30167 Hannover, Germany} \\[4pt]
{Email: lechtenf@itp.uni-hannover.de}
\vskip 0.5cm

\end{center}
\vskip 1cm
\begin{abstract} \noindent
Recent years have seen an upsurge of interest in dynamical realizations
of the superconformal group SU$(1,1|2)$ in mechanics.
Remarking that SU$(1,1|2)$ is a particular member of a chain
of supergroups SU$(1,1|n)$ parametrized by an integer~$n$,
here we begin a systematic study of SU$(1,1|n)$ multi-particle mechanics.
A representation of the superconformal algebra $su(1,1|n)$ is constructed
on the phase space spanned by $m$~copies of the $(1,2n,2n{-}1)$ supermultiplet.
We show that the dynamics is governed by two prepotentials $V$ and~$F$,
and the Witten-Dijkgraaf-Verlinde-Verlinde equation for~$F$ shows up
as a consequence of a more general fourth-order equation. All solutions to the latter
in terms of root systems reveal decoupled models only. An extension
of the dynamical content of the $(1,2n,2n{-}1)$ supermultiplet by angular
variables in a way similar to the SU$(1,1|2)$ case is problematic.
\end{abstract}

\vspace{0.5cm}

\end{titlepage}

\renewcommand{\thefootnote}{\arabic{footnote}}
\setcounter{footnote}0

\noindent
{\bf 1. Introduction}\\

\noindent
The recent increase of interest in dynamical realizations of the superconformal group SU$(1,1|2)$ \cite{W}--\cite{AG} and its $D(2,1|\alpha)$ extension \cite{IKL}--\cite{IF} was motivated by the proposal in
\cite{claus,gibb} that a study of superconformal mechanics may have applications to the quantum mechanics of black holes. In particular, according to \cite{gibb} the large-$m$ limit of the $m$-particle SU$(1,1|2)$ superconformal Calogero model may provide a microscopic description of the extreme Reissner--Nordstr{\"o}m black hole in the near-horizon limit.

The explicit construction of the $m$-particle SU$(1,1|2)$ superconformal Calogero model reduces to solving a variant of the Witten--Dijkgraaf--Verlinde--Verlinde (WDVV) equation \cite{W,bgl}. Although plenty of interesting solutions
to the WDVV equation were found in terms of root systems and their deformations \cite{W,GLP,GLP1,MG,LP,LST,veselov}, the construction of interacting models seems unfeasible beyond $m=3$.
Since, in the context of~\cite{gibb}, it is the structure of the superconformal group which matters, any multi-particle SU$(1,1|2)$
mechanics appears to be a good candidate. Yet, no attempt has been made to link the large-$m$ limit of any known superconformal many-body quantum mechanics to the extreme Reissner--Nordstr{\"o}m black hole in the near-horizon limit.

The studies in \cite{W}--\cite{IF} proved useful for understanding the structure of interactions of various SU$(1,1|2)$ and $D(2,1|\alpha)$ supermultiplets. Supersymmetric couplings in $d{=}1$ are of interest on their own right
because of novel features which are absent in higher dimensions.

The superconformal group SU$(1,1|2)$ is a particular member of a chain of supergroups SU$(1,1|n)$ parametrized by an integer~$n$. The corresponding superconformal algebra $su(1,1|n)$ involves $n^2{+}3$ bosonic and $4n$ fermionic generators. In particular, its bosonic sector includes $so(2,1)$ and $su(n)$ subalgebras.
The natural question arises whether the interesting features revealed for the SU$(1,1|2)$ case persist for higher values of $n$. Interacting many-body SU$(1,1|n)$ mechanics may also have applications to the quantum mechanics of higher-dimensional black holes. Specifically, the bosonic subgroup SO$(2,1)\times SU(n)$ of SU$(1,1|n)$ coincides with the near-horizon symmetry group of the  Myers--Perry black hole with all rotation parameters set equal (see, e.g., the discussion in \cite{G3,GNS}).

In this work, we initiate a systematic study of SU$(1,1|n)$ many-body mechanics. There are two competing approaches to analyzing superconformal mechanics, namely the direct construction of an $su(1,1|n)$ representation within the Hamiltonian framework, and the superfield approach combined with the method of nonlinear realizations.
In~\cite{IvKrLe} for example, the second approach has been used to describe a single supermultiplet of  type $(1,2n,2n{-}1)$.
Although the superfield formulation is more powerful, the Hamiltonian approach yields on-shell components and allows one to comprehend the basic dynamical features and the structure of interactions in a simpler and more transparent way.
In some instances it also offers notable technical simplifications in building interacting models~\cite{AG}. In this work we thus adhere to the Hamiltonian formalism.

The paper is organized as follows. In Section~2 we fix our notation and represent the structure relations of the superconformal algebra $su(1,1|n)$
in a form analogous to the previously studied case of $su(1,1|2)$. Section~3 is devoted to the construction of an $su(1,1|n)$ representation on the phase space spanned by $m$~copies of the $(1,2n,2n-1)$ supermultiplet. It is shown that similarly to the SU$(1,1|2)$ case the dynamics is governed by two prepotentials $V$ and~$F$. However, the WDVV equation appears as a consequence of a more general fourth-order equation for $F$. The latter is absent in SU$(1,1|2)$ mechanics because of a specific Fierz identity which exists for SU$(2)$ spinors only. In Section~4 we consider
prepotentials~$F$ constructed from root systems. It is demonstrated that the fourth-order structure equation characterizing SU$(1,1|n)$ mechanics forces all root vectors to be mutually orthogonal. This implies decoupled dynamics. In Section~5 we try to generalize also the analysis of~\cite{AG} from SU$(1,1|2)$ to SU$(1,1|n)$. More specifically, we attempt to extend an arbitrary phase-space representation of~$su(n)$ to an $su(1,1|n)$ representation, with a negative result. The concluding Section~6 contains a summary and an outlook. Throughout the paper summation over repeated indices is understood.

\vspace{0.5cm}

\noindent
{\bf 2. Superconformal algebra $su(1,1|n)$}\\

\noindent
The superconformal algebra $su(1,1|n)$ involves $n^2+3$ bosonic and $4n$ fermionic generators. Its even part is the direct sum $so(2,1)\oplus su(n)\oplus u(1)$. The generators of $so(2,1)$, which we designate as $H$, $D$, $K$, correspond to the time translation, dilatation and special conformal transformation, respectively. The $R$--symmetry subalgebra $su(n)\oplus u(1)$ is generated by $J_a$, with $a=1,\dots,n^2-1$, and
$L$. The odd part of the superalgebra includes the supersymmetry generators $Q_\alpha$, $\bar Q^{\alpha}$, where $\alpha=1,\dots,n$, and their superconformal partners $S_\alpha$, $\bar S^{\alpha}$. It is assumed that the fermions are hermitian conjugates of each other
\be
{(Q_\alpha)}^{\dagger}=\bar Q^{\a}, \qquad  {(S_\alpha)}^{\dagger}=\bar S^{\a}.
\ee
$Q_\alpha$ and $S_\alpha$ transform as $su(n)$ spinors.
The structure relations of $su(1,1|n)$ read
\begin{align}\label{algebra}
&
\{ H,D \}=H, && \{ H,K \}=2D,
\nonumber\\[2pt]
&
\{D,K\}=K, && \{ J_a,J_b \}=f_{abc} J_c,
\nonumber\\[2pt]
&
\{ Q_\a, \bar Q^\b \}=-2 i H {\d_\a}^\b, &&
\{ Q_\a, \bar S^\b \}=2{{(\lambda_a)}_\a}^\b J_a+\left(2iD-\sfrac{n{-}2}{n} L\right){\d_\a}^\b,
\nonumber\\[2pt]
&
\{ S_\a, \bar S^\b \}=-2i K {\d_\a}^\b, &&
\{ \bar Q^\a, S_\b \}=-2{{(\lambda_a)}_\b}^\a J_a+\left(2iD+\sfrac{n{-}2}{n} L\right) {\d_\b}^\a,
\nonumber\\[2pt]
& \{ D,Q_\a\} = -\sfrac{1}{2} Q_\a, && \{ D,S_\a\} =\sfrac{1}{2} S_\a,
\nonumber\\[2pt]
&
\{ K,Q_\a \} =S_\a, && \{ H,S_\a \}=-Q_\a,
\nonumber\\[2pt]
&
\{ J_a,Q_\a\} =\sfrac{i}{2} {{(\lambda_a)}_\a}^\b Q_\b, && \{ J_a,S_\a\} =\sfrac{i}{2} {{(\lambda_a)}_\a}^\b S_\b,
\nonumber\\[2pt]
& \{ D,\bar Q^\a \} =-\sfrac{1}{2} \bar Q^\a, && \{ D,\bar S^\a\} =\sfrac{1}{2} \bar S^\a,
\nonumber\\[2pt]
& \{K,\bar Q^\a\} =\bar S^\a, && \{ H,\bar S^\a\} =-\bar Q^\a,
\nonumber\\[2pt]
&
\{J_a,\bar Q^\a\} =-\sfrac{i}{2} \bar Q^\b {{(\lambda_a)}_\b}^\a, && \{ J_a,\bar S^\a\} =-\sfrac{i}{2}
\bar S^\b {{(\lambda_a)}_\b}^\a,
\nonumber\\[2pt]
&
\{L,Q_\a\} =i Q_\a , && \{L,S_\a\} =i S_\a,
\nonumber\\[2pt]
&
\{L,\bar Q^\a\} =-i \bar Q^\a , && \{L,\bar S^\a\} =-i \bar S^\a ,
\end{align}
where $f_{abc}$ are the totally antisymmetric structure constants of $su(n)$ and $\lambda_a$ are the hermitian and traceless $n\times n$--matrices which obey the (anti)commutation relations
\bea\label{lm}
[\lambda_a,\lambda_b]=2 i f_{abc} \lambda_c, \qquad
\{\lambda_a,\lambda_b \}=\sfrac 43 \delta_{ab}+2 d_{abc} \lambda_c,
\eea
with the totally symmetric coefficients $d_{abc}$.
In what follows the Fierz identity
\be\label{FI}
\frac 12 {{(\lambda_a)}_\a}^\b {{(\lambda_a)}_\gamma}^\sigma=-\frac{1}{n} {\delta_\alpha}^\beta  {\delta_\gamma}^\sigma+  {\delta_\gamma}^\beta  {\delta_\alpha}^\sigma
\ee
proves to be helpful.

\vspace{0.5cm}

\noindent
{\bf 3. Realization of $su(1,1|n)$ in many-body mechanics}\\

\noindent
In order to realize the $su(1,1|n)$ superconformal algebra in many-body mechanics, let us consider a phase space parametrized by $m$ bosonic canonical pairs $(x^i, p^i)$, and $m$ self--conjugate fermions  ${(\psi^i_\alpha)}^{\dagger}=\bar\p^{ i \a}$, $i=1,\dots,m$, $\alpha=1,\dots,n$, which obey
the conventional Poisson brackets
\be\label{cr}
\{x^i,p^j \}=\delta^{ij} , \qquad \{ \p^i_\a, \bar\p^{j \b} \}=-i{\d_\a}^\b \delta^{ij}.
\ee
It is assumed that each fermion belongs to the fundamental representation of SU$(n)$.

Guided by the previous studies of the $su(1,1|2)$--case \cite{W,bgl,GLP}, let us introduce two prepotentials $V(x^1,\dots,x^n)$, $F(x^1,\dots,x^n)$ and consider the following functions:
\begin{align}\label{repr}
&
H=\sfrac{1}{2} \left(p^i p^i+\partial^i V \partial^i V  \right)+\partial^i \partial^j V (\bar\psi^i \psi^j)+\sfrac 12 \partial^i \partial^j \partial^k \partial^l F  (\bar\psi^i \psi^j) (\bar\psi^k \psi^l), && D=tH-\sfrac 12 x^i p^i,
\nonumber\\[2pt]
&
K=t^2 H-t x^i p^i +\sfrac 12 x^i x^i, && J_a=\sfrac 12 (\bar\p^i \lambda_a \p^i),
\nonumber\\[2pt]
&
Q_\a=(p^i+i \partial^i V) \p^i_\a+i \partial^i \partial^j \partial^k F \p^i_\a (\bar\psi^j \psi^k) , && S_\a=x^i \p^i_\a -t Q_\a,
\nonumber\\[2pt]
&
\bar Q^\a =(p^i-i \partial^i V) \bar\p^{i \a}-i \partial^i \partial^j \partial^k F \bar\p^{i \a} (\bar\psi^j \psi^k), &&
\bar S^\a=x^i \bar\p^{i \a} -t \bar Q^\a,
\nonumber\\[2pt]
&
L=\bar\psi^i \psi^i,
\end{align}
where $\bar\psi^i \psi^j=\bar\psi^{i \alpha} \psi^j_\alpha$, $\bar\p^i \lambda_a \p^i=\bar\p^{i \alpha} {{(\lambda_a)}_\alpha}^\beta \p_\beta^i$. It is straightforward to verify that these functions do obey the structure relations (\ref{algebra}) under the Poisson bracket (\ref{cr}) provided the restrictions
on the prepotentials
\bea\label{con}
&&
(\partial^i \partial^j \partial^k F)( \partial^k \partial^l \partial^m F)\=(\partial^m \partial^j \partial^k F)( \partial^k \partial^l \partial^i F),  \qquad x^i (\partial^i \partial^j \partial^k F)\=-\delta^{jk},
\\[4pt]
&&
x^i \partial^i V=C, \qquad \partial^i \partial^j V\= (\partial^i \partial^j \partial^k F) \partial^k V, \qquad
\partial^i \partial^j \partial^k \partial^l F\=(\partial^i \partial^j \partial^p F)(\partial^p \partial^k \partial^l F).
\nonumber
\eea
hold, with $C$ being an arbitrary constant.

Note that all the constraints in (\ref{con}) coincide with those characterizing the $su(1,1|2)$ case, but
for the rightmost equation entering the second line which is new. It arises
when computing the bracket $\{ Q_\a, \bar Q^\b \}$ which explicitly involves the term
\be
\bigl(\partial^i \partial^j \partial^k \partial^l F-(\partial^i \partial^j \partial^p F)(\partial^p \partial^k \partial^l F)\bigr)\,\p^i_\alpha \bar\p^{j \beta} (\bar\psi^k \psi^l).
\ee
For $n=2$ the spinor index $\alpha$ takes only two values, and the spinors in the previous formula can be reordered so as to yield the piece proportional to $(\bar\psi^i \psi^j)(\bar\psi^k \psi^l) {\delta_\alpha}^\beta$, thus providing a contribution to the Hamiltonian which is quartic in fermions. For $n>2$ such reordering is no longer possible, and one has to impose the extra condition
\be\label{extra}
\partial^i \partial^j \partial^k \partial^l F\=(\partial^i \partial^j \partial^p F)(\partial^p \partial^k \partial^l F),
\ee
which yields the main difference from the $su(1,1|2)$ case. Note that, by antisymmetrization of the indices $i$ and $l$, (\ref{extra}) actually implies the WDVV equation visible in the first line in~(\ref{con}). Hence, the additional requirement as compared to the $n=2$ case is the totally symmetric projection of (\ref{extra}),
\be\label{extrasym}
\partial^i \partial^j \partial^k \partial^l F \=
(\partial^{(i} \partial^j \partial^p F)(\partial^p \partial^k \partial^{l)} F) ,
\ee
where the symmetrisation (with weight $\frac1{4!}$) excludes the summation index~$p$.
Further differentiation of this relation, together with the WDVV equation, yields a hierarchy of equations,
\be
\partial^{i_1}\partial^{i_2}\cdots\partial^{i_{r+3}} F \= r!
(\partial^{i_1}\partial^{i_2}\partial^{k_1} F)
(\partial^{k_1}\partial^{i_3}\partial^{k_2} F)
(\partial^{k_2}\partial^{i_4}\partial^{k_3} F) \cdots
(\partial ^{k_r}\partial^{i_{r+2}}\partial^{i_{r+3}} F)
\ee
together with
$x^i(\partial^i\partial^{i_2}\cdots\partial^{i_{r+3}} F)=-r\,\partial^{i_2}\cdots\partial^{i_{r+3}} F$,
for $r=1,2,\ldots.$

When computing the brackets
$\{ Q_\a, \bar S^\b \}$ and $\{ \bar Q^\a, S_\b \}$, one has to use the Fierz identity~(\ref{FI}).
In particular, the constant $C$, which enters the homogeneity condition $x^i \partial^i V=C$, appears in the algebra as the central charge,
\bea\label{strr}
&&
\{ Q_\a, \bar S^\b \}\=\phantom{-}2{{(\lambda_a)}_\a}^\b J_a+\left(2iD-\sfrac{n{-}2}{n} L+C\right){\d_\a}^\b,
\nonumber\\[4pt]
&&
\{ \bar Q^\a, S_\b \}\=-2{{(\lambda_a)}_\b}^\a J_a+\left(2iD+\sfrac{n{-}2}{n} L-C\right) {\d_\b}^\a.
\eea
If desirable, $C$ can be removed by redefining $L$. In the latter case the bosonic limit of $L$ yields a constant rather than zero.

\vspace{0.5cm}

\noindent
{\bf 4. Prepotentials $F$ related to root systems}\\

\noindent
The leftmost equation in the first line in (\ref{con}) is a variant of the WDVV equation.
With regard to the SU($1,1|2)$ mechanics it has been extensively studied
in \cite{bgl,GLP,GLP1,LP,LST}. In particular, each solution of the WDVV equation
satisfying (\ref{extrasym}) will qualify to describe some SU$(1,1|n)$ superconformal mechanics.
The known WDVV solutions are based on so-called $\vee$-systems~\cite{veselov},
which are certain deformations of Coxeter root systems.
For these, the prepotential~$F$ takes the form
\be\label{Fansatz}
F \= -\sfrac14 \sum_\a h_\a\, (\a\cdot x)^2 \ln(\a\cdot x)^2 ,
\ee
where $\{\a\}$ is a set of positive $m$-dimensional root vectors,
subject to the usual constraints for reflection groups or their $\vee$-system deformations,
and $h_\a$ are real weights to be determined.
Inserting (\ref{Fansatz}) into (\ref{extra}), we obtain the condition
\be
\sum_\a h_\a \, \frac{\a^i\,\a^j\,\a^k\,\a^l}{(\a\cdot x)^2} \ +\
\sum_{\a,\b} h_\a h_\b\, \frac{\a^i\,\a^j\,(\a\cdot\b)\,\b^k\,\b^l}{(\a\cdot x)(\b\cdot x)} \= 0 .
\ee
The diagonal terms in this double sum fix the weights,
\be
(\a\cdot\a)\,h_\a = 1.
\ee
The projection antisymmetric in $i$ and $l$ ensures the WDVV equation;
it is assumed to be fulfilled for our root systems.
The symmetric projection gives further algebraic conditions:
the vanishing of the double residues of the poles
$(\a\cdot x)^{-1}(\b\cdot x)^{-1}$ for any pair $(\a,\b)$ yields
\be
(\a\cdot\b)\,(\a^i \a^j \b^k \b^l + \b^i \b^j \a^k \a^l) \= 0 .
\ee
Contracting this with $\a^i \b^j \a^k \b^l$ produces
\be
(\a\cdot\a) (\b\cdot\b) (\a\cdot\b)^2 \= 0
\qquad\Longrightarrow\qquad \a\cdot\b = 0
\ee
for any pair of distinct roots $(\a,\b)$.
This admits only the direct sum of mutually orthogonal one-dimensional (i.e.\ rank-one) systems.
By a rigid rotation of coordinates $x^i$, one can always bring it into the form
\be
\{\a\} \= \{ (1,0,0,\ldots,0), (0,1,0,\ldots,0), \ldots, (0,0,0,\ldots,1) \} .
\ee
So we have arrived at a no-go theorem for interacting SU$(1,1|n)$ mechanics based on
the on-shell supermultiplet of type~$(1,2n,2n{-}1)$.

\vspace{0.5cm}

\noindent
{\bf 5. Angular variables}\\

\noindent
For SU$(1,1|2)$ mechanics one can extend the dynamical content of the simplest
$(1,4,3)$ supermultiplet by introducing angular variables providing some realization of $su(2)$
in a purely group-theoretic way \cite{g1,AG}. It suffices to consider
a phase space parametrized by the canonical pairs $(\theta^A, p_{\theta A})$, $A=1,\dots,n$,
which obey the conventional Poisson brackets
\be
\{\theta^A,p_{\theta B} \}={\delta^A}_B
\ee
and realize on such a phase space the functions $J_a=J_a (\theta,p_\theta)$, $a=1,2,3$, which obey the structure relations of the $su(2)$ $R$-symmetry subalgebra
\be\label{su(2)}
\{J_a,J_b \}=\e_{abc} J_c.
\ee
Then, the supersymmetry charges involve the angular variables only via the currents $J_a$,
\bea
&&
Q_\a\=p \,\psi_\a+\frac{2i}{x} {(\s_a \psi)}_\alpha J_a -\frac{i}{x} \psi_\a (\bar\psi \psi),
\eea
where ${{(\sigma_a)}_\alpha}^\beta$ are the Pauli matrices \cite{AG}.

Let us try to generalize this construction to the case of the superconformal algebra $su(1,1|n)$.
Introducing functions $J_ a$ of the angular variables subject to the $su(n)$ structure relations
$\{ J_a,J_b \}=f_{abc} J_c$, and employing the matrices ${{(\lambda_a)}_\alpha}^\beta$ from~(\ref{lm}),
it is straightforward to verify that the obvious candidate supersymmetry charge,
\bea
&&
Q_\a\=p \,\psi_\a+\frac{2i}{x} {(\lambda_a \psi)}_\alpha  J_a -\frac{i}{x} \psi_\a (\bar\psi \psi),
\eea
is indeed nilpotent, i.e.
\be
\{Q_\a,Q_\b \}=0.
\ee
However, in view of the properties of the $\lambda$--matrices in (\ref{lm}), the bracket of $Q_\a$ with $\bar Q^\beta$ yields not just the Hamiltonian:
\be
\{Q_\a, \bar Q^\beta\}\=-2 i\,H {\delta_\a}^\beta-\frac{4 i}{x^2} {{(\lambda_a)}_\alpha}^\beta d_{abc} J_b J_c,
\ee
where $d_{abc}$ are the symmetric structure coefficients appearing in (\ref{lm}).
This means that the algebra does not close. One might try to modify the troublesome second term in $Q_\a$. However, such a term seems indispensable for providing the structure relations~(\ref{strr}).

We thus conclude that an extension of the $(1,2n,2n{-}1)$ supermultiplet by angular variables in a way similar to the $su(1,1|2)$ case is problematic. Perhaps a more sophisticated construction involving extra auxiliary variables will help to circumvent the problem.

\vspace{0.5cm}

\noindent
{\bf 6. Discussion}\\

\noindent
To summarize, in this work we made the first step towards a systematic description of SU$(1,1|n)$ multi-particle superconformal mechanics. Our consideration was primarily focused on the possibilities offered by the Hamiltonian formalism. The structure relations of the superconformal algebra $su(1,1|n)$ were established in a form analogous to the previously studied $su(1,1|2)$ case. A representation of $su(1,1|n)$ on the phase space spanned by $m$~copies of the $(1,2n,2n{-}1)$ supermultiplet was constructed. It was shown that the dynamics is governed by two prepotentials $V$ and~$F$,
and that the WDVV equation for~$F$ arises as a consequence of a more restrictive fourth-order equation. Solutions to the latter in terms of root systems allow decoupled models only.  An attempt to extend the dynamical content of the $(1,2n,2n{-}1)$ supermultiplet by adding angular variables in a way similar to the $su(1,1|2)$ case compromised the closure of the $su(1,1|n)$ superconformal algebra. Hence, our results indicate that the construction of interacting SU$(1,1|n)$ models with $n>2$ appears to be a more difficult task than in the SU$(1,1|2)$ case.

The Hamiltonian formulation adopted in this work automatically yields on-shell models.
It is tempting to investigate SU$(1,1|n)$ mechanics off-shell within the superfield approach combined with the method of nonlinear realizations, along the lines proposed in~\cite{IvKrLe}.
The key problem within the superfield method will be to guess the superfield constraints which will result in interacting dynamics. A possible link of SU$(1,1|n)$ mechanics to the near-horizon Myers--Perry black hole with equal rotation parameters is worth studying as well.
Finally, it might be rewarding to investigate the integrability of~(\ref{extra}) on its own, which is more special than the WDVV~equation.

\vskip 0.5cm

\noindent
{\bf Acknowledgements}\\

\noindent
We thank S.~Krivonos for bringing \cite{IvKrLe} to our attention.
A.G. is grateful to the Institute for Theoretical Physics at Hannover University for the hospitality extended to him at different stages of this research.
The work was supported by the DFG grant Le-838/12-2, the MSE program Nauka under the project 3.825.2014/K, the RFBR grant 15-52-05022, and the
Action MP1405 QSPACE from the European Cooperation in Science and Technology (COST).


\begin{thebibliography}{nn}
\addtolength{\itemsep}{-7pt}

\bibitem{W}
N. Wyllard, {\it (Super)conformal many body quantum mechanics with extended supersymmetry}, J. Math. Phys. {\bf 41} (2000) 2826, hep-th/9910160.
\bibitem{Gal}
A. Galajinsky, {\it Remarks on N{=}4 superconformal extension of the Calogero model}, \\
Mod. Phys. Lett. A {\bf 18} (2003) 1493, hep-th/0302156.
\bibitem{BGK}
S. Bellucci, A. Galajinsky, S. Krivonos, {\it New many-body superconformal models as reductions of simple composite systems}, Phys. Rev. D {\bf 68} (2003) 064010, hep-th/0304087.
\bibitem{bgl}
S. Bellucci, A. Galajinsky, E. Latini, {\it New insight into WDVV equation}, \\
Phys. Rev. D {\bf 71} (2005) 044023, hep-th/0411232.
\bibitem{DI1}
F. Delduc, E. Ivanov, {\it Gauging N=4 supersymmetric mechanics II: (1,4,3) models from the (4,4,0) ones}, Nucl. Phys. B {\bf 770} (2007) 179, hep-th/0611247.
\bibitem{GLP}
A. Galajinsky, O. Lechtenfeld, K. Polovnikov, {\it N=4 superconformal Calogero models},\\
 JHEP {\bf 0711} (2007) 008, arXiv:0708.1075.
\bibitem{GLP1}
A. Galajinsky, O. Lechtenfeld, K. Polovnikov, {\it N=4 mechanics, WDVV equations and roots}, JHEP {\bf 0903} (2009) 113, arXiv:0802.4386.
\bibitem{bks}
S. Bellucci, S. Krivonos, A. Sutulin, {\it N=4 supersymmetric 3-particles Calogero model}, \\
Nucl. Phys. B {\bf 805} (2008), arXiv:0805.3480.
\bibitem{g1}
A. Galajinsky, {\it  	
Particle dynamics on ${AdS}_2 \times {S}^2$ background with two-form flux}, \\
Phys. Rev. D {\bf 78} (2008) 044014, arXiv:0806.1629.
\bibitem{FIL}
S. Fedoruk, E. Ivanov, O. Lechtenfeld, {\it Supersymmetric Calogero models by gauging}, \\
Phys. Rev. D {\bf 79} (2009) 105015, arXiv:0812.4276.
\bibitem{KLP}
S. Krivonos, O. Lechtenfeld, K. Polovnikov, {\it  	
N=4 superconformal n-particle mechanics via superspace}, Nucl. Phys. B {\bf 817} (2009) 265, arXiv:0812.5062.
\bibitem{IFL1}
S. Fedoruk, E. Ivanov, O. Lechtenfeld, {\it Superconformal mechanics}, \\
J. Phys. A {\bf 45} (2012) 173001, arXiv:1112.1947.
\bibitem{CSJ}
N.B. Copland, Sung Moon Ko, Jeong-Hyuck Park, 
{\it Superconformal Yang-Mills quantum mechanics and Calogero model with OSp$(N|2,R)$ symmetry}, \\
JHEP {\bf 1207} (2012) 076, arXiv:1205.3869. 
\bibitem{AG}
A. Galajinsky, {\it  N=4 superconformal mechanics from the $su(2)$ perspective}, \\
JHEP {\bf 1502} (2015) 091, arXiv:1412.4467.
\bibitem{IKL}
E. Ivanov, S. Krivonos, O. Lechtenfeld, {\it New variant of N=4 superconformal mechanics}, \\
JHEP {\bf 03} (2003) 014, hep-th/0212303.
\bibitem{IL2}
E. Ivanov, O. Lechtenfeld, {\it  	
N=4 supersymmetric mechanics in harmonic superspace}, \\
JHEP {\bf 0309} (2003) 073, hep-th/0307111.
\bibitem{IKL4}
E. Ivanov, S. Krivonos, O. Lechtenfeld, {\it
$N=4$, $d = 1$ supermultiplets from nonlinear realizations of $D(2,1;\alpha)$}, Class. Quant. Grav. {\bf 21} (2004) 1031, hep-th/0310299.
\bibitem{BK2}
S. Bellucci, S. Krivonos, {\it  	
Potentials in N=4 superconformal mechanics}, \\
Phys. Rev. D {\bf 80} (2009) 065022, arXiv:0905.4633.
\bibitem{HKLN}
T. Hakobyan, S. Krivonos, O. Lechtenfeld, A. Nersessian, {\it Hidden symmetries of integrable conformal mechanical systems}, Phys. Lett. A374 (2010) 801, arXiv:0908.3290.
\bibitem{FIL2}
S. Fedoruk, E. Ivanov, O. Lechtenfeld, {\it  	
New $D(2,1;\alpha)$ mechanics with spin variables}, \\
JHEP {\bf 1004} (2010) 129, arXiv:0912.3508.
\bibitem{KL1}
S. Krivonos, O. Lechtenfeld, {\it Many-particle mechanics with $D(2,1;\alpha)$ superconformal symmetry}, JHEP {\bf 1102} (2011) 042, arXiv:1012.4639.
\bibitem{GG}
K. Govil, M. G\"unaydin, {\it Minimal unitary representation of $D(2,1;\lambda)$ and its SU$(2)$ deformations and $d{=}1$, $N{=}4$ superconformal models}, \\
Nucl. Phys. B {\bf 869} (2013) 111, arXiv:1209.0233.
\bibitem{IF}
S. Fedoruk, E. Ivanov, {\it  New realizations of the supergroup $D(2,1;\alpha)$ in $N{=}4$ superconformal mechanics}, JHEP {\bf 1510} (2015) 087, arXiv:1507.08584.
\bibitem{claus}
P. Claus, M. Derix, R. Kallosh, J. Kumar,
P.K. Townsend, A. Van Proeyen, {\it Black holes and superconformal mechanics},
Phys. Rev. Lett. {\bf 81} (1998) 4553, hep-th/9804177.
\bibitem{gibb}
G.W. Gibbons, P.K. Townsend, {\it Black holes and Calogero models},\\
Phys. Lett. B {\bf 454} (1999) 187, hep-th/9812034.
\bibitem{MG}
R. Martini, P.K.H. Gragert, {\it Solutions of WDVV equations in Seiberg-Witten theory from root systems}, J. Nonlin. Math. Phys. {\bf 6} (1999) 1, hep-th/9901166.
\bibitem{LP} 	
O. Lechtenfeld, K. Polovnikov, {\it  A new class of solutions to the WDVV equation}, Phys. Lett. A {\bf 374} (2010) 504, arXiv:0907.2244.
\bibitem{LST}
O. Lechtenfeld, K. Schwerdtfeger, J. Th\"urigen, {\it $N{=}4$ multi-particle mechanics, WDVV equation and roots},
SIGMA {\bf 7} (2011) 023, arXiv:1011.2207.
\bibitem{veselov}
A.P. Veselov, {\it Deformations of the root systems and new solutions to generalised WDVV equations},
Phys. Lett. A {\bf 261} (1999), 297, hep-th/9902142.
\bibitem{G3}
A. Galajinsky, {\it Near horizon black holes in diverse dimensions and integrable models}, \\
Phys. Rev. D {\bf 87} (2013) 024023, arXiv:1209.5034.
\bibitem{GNS}
A. Galajinsky, A. Nersessian, A. Saghatelian, {\it Superintegrable models related to near horizon extremal Myers-Perry black hole in arbitrary dimension}, \\
JHEP {\bf 1306} (2013) 002, arXiv:1303.4901.
\bibitem{IvKrLe}
E.A. Ivanov, S.O. Krivonos, V.M. Leviant, {\it Geometric superfield approach to superconformal mechanics},  J. Phys. A {\bf 22} (1989) 4201.

\end{thebibliography}
\end{document}